\documentclass[amsmath,amssymb,aps,pra,twocolumn,superscriptaddress]{revtex4-1}

\usepackage{graphicx}% Include figure files
\usepackage{dcolumn}% Align table columns on decimal point
\usepackage{bm}% bold math
\usepackage{enumitem}
\usepackage{textcomp}
\usepackage{gensymb}
\begin{document}

%\graphicspath{{figures/emf_jpg/}}

\preprint{CEI_CO_Seq_Conc}

\title{Investigation of the Carbon Monoxide Dication Lifetime Using (CO)$_2$ Dimer Fragmentation}

\author{A. M\'ery} \affiliation{CIMAP, CEA-CNRS-ENSICAEN-UNICAEN, Normandie Universit\'e,\\ BP5133, F-14050 Caen Cedex 04, France}
\author{X. Fl\'echard}
\email{flechard@lpccaen.in2p3.fr}
\affiliation{Normandie Univ, ENSICAEN, UNICAEN, CNRS/IN2P3, LPC Caen, 14000 Caen, France}
\author{S. Guillous} \affiliation{CIMAP, CEA-CNRS-ENSICAEN-UNICAEN, Normandie Universit\'e,\\ BP5133, F-14050 Caen Cedex 04, France}
\author{V. Kumar} \affiliation{CIMAP, CEA-CNRS-ENSICAEN-UNICAEN, Normandie Universit\'e,\\ BP5133, F-14050 Caen Cedex 04, France}
\affiliation{Inter University Accelerator Center, Aruna Asaf Ali Marg, New Delhi, Delhi 110067}
\author{M. Lalande} \affiliation{CIMAP, CEA-CNRS-ENSICAEN-UNICAEN, Normandie Universit\'e,\\ BP5133, F-14050 Caen Cedex 04, France}
\author{J. Rangama} \affiliation{CIMAP, CEA-CNRS-ENSICAEN-UNICAEN, Normandie Universit\'e,\\ BP5133, F-14050 Caen Cedex 04, France}
\author{W. Wolff} \affiliation{Instituto de F\'isica - Universidade Federal do Rio de Janeiro, Cidade Universit\'aria, Rio de Janeiro, Brazil}
\author{A. Cassimi} \affiliation{CIMAP, CEA-CNRS-ENSICAEN-UNICAEN, Normandie Universit\'e,\\ BP5133, F-14050 Caen Cedex 04, France}

\date{\today}

\begin{abstract}
The fragmentation of carbon monoxide dimers induced by collisions with low energy Ar$^{9+}$ ions is investigated using the COLTRIMS technique. The presence of a neighbor molecule in the dimer serves here as a diagnostic tool to probe the lifetimes of the $\rm CO^{2+}$ molecular dications resulting from the collision. The existence of metastable states with lifetimes ranging from 2~ps to 200~ns is clearly evidenced experimentally through a sequential 3-body fragmentation of the dimer, whereas fast dissociation channels are observed in a so-called concerted 3-body fragmentation process.
The fast fragmentation process leads to a kinetic energy release distribution also observed in collisions with monomer CO targets. This is found in contradiction with the conclusions of a former study attributing this fast process to the perturbation induced by the neighbor molecular ion.
\end{abstract}



\maketitle
%%%%%%%%%%%%%%%%%%%%%%%%%%%%%%%%%%%%%%%%%%%%
%%%%%%%%%%%%%%%%%%%%%%%%%%%%%%%%%%%%%%%%%%%%
%%%%%%%%%%%%%%%%%%%%%%%%%%%%%%%%%%%%%%%%%%%%
\section{INTRODUCTION}
Doubly charged diatomic molecular ions have unusual properties leading to a wide range of lifetimes against dissociation. Their stability depends on both the accessible decay mechanisms and the position of the populated rovibronic levels with respect to the barrier height. Among the large variety of doubly charged diatomic species investigated so far, CO$^{2+}$ is probably the one that has attracted the largest interest, with measured lifetimes ranging from submicroseconds to a few seconds.
Since its first observation within a mass spectrometer in 1932 [\onlinecite{Friedlander1932}], the lifetime of metastable states of $\rm CO^{2+}$ has been widely studied both experimentally and theoretically. The existence of excited states with lifetimes in the millisecond range and up to few seconds have been demonstrated using an ion storage ring [\onlinecite{Andersen1993}]. Shorter lifetimes in the 10 ns to 1~$\rm \mu$s range have also been identified [\onlinecite{PenentPRL1998},\onlinecite{BouhnikPRA2001}]. In parallel, computational studies have shown that only the lowest vibronic levels ($\rm \upsilon \leq 3$) of the electronic ground state ($\rm ^3\Pi$) and low lying states ($\rm ^1\Sigma^+$, $\rm ^1\Pi$ and $\rm ^3\Sigma^+$) of $\rm CO^{2+}$ have lifetimes longer than 1~ns [\onlinecite{WetmoreJCP1984},\onlinecite{MrugalaJCP2008},\onlinecite{Sedivcova2006}]. The decay rate of these states was found to be mainly governed by tunneling or by predissociation to the repulsive $\rm ^3\Sigma^-$ state [\onlinecite{PandeyJCP2014}].

More recently, a fast dissociation channel of the $\rm CO^{2+}$ molecular dication has been observed in the 3-body breakup of triply charged dimers $\rm (CO)_2^{3+}$ ionized by intense ultrashort laser pulses [\onlinecite{DingPRL2017}]. However, this fast process was found to be associated with a higher kinetic energy release (KER) distribution than expected when compared to the dissociation of the isolated $\rm CO^{2+}$ molecular ion. The authors suggested that this fast process would only exist in a van der Waals complex thanks to the weak coupling with a neighbor $\rm CO^{+}$ molecular ion. The symmetry breaking induced by the counter charge then leads to an avoided crossing with a dissociation channel, whereas the monomer would dissociate only via a weaker spin-orbit coupling.  

In the present work, we revisit this investigation using Coulomb explosion imaging (CEI) of triply charged (CO)$_2^{3+}$ dimers ionized by low energy Ar$^{9+}$ ions. The presence of a neighbor molecular CO$^+$ ion serves here as a probe to get insight into the dissociation process of the other CO$^{2+}$ molecular ion resulting from the collision. The detailed study of the kinematic of the 3-body fragmentation channel provides a clear signature of the lifetime ranges of the CO$^{2+}$ dication for several sets of data. The corresponding KER distributions are then used to identify the states of the  $\rm CO^{2+}$ molecular ion initially populated by the collision process. This new investigation differs from the work of [\onlinecite{DingPRL2017}] by two aspects: i) the use of low energy highly charged ions as projectiles, which leads to multiple electron capture from the target and to the population of states with higher excitation energy, and ii) a higher momentum resolution, which allows a clearer comparison with the dissociation of isolated $\rm CO^{2+}$ molecular ions. 
%%%%%%%%%%%%%%%%%%%%%%%%%%%%%%%%%%%%%%%%%%%%
%%%%%%%%%%%%%%%%%%%%%%%%%%%%%%%%%%%%%%%%%%%%
%%%%%%%%%%%%%%%%%%%%%%%%%%%%%%%%%%%%%%%%%%%%
\section{EXPERIMENT}
The cold target recoil ion momentum spectroscopy (COLTRIMS) setup used for this work has already been described in detail elsewhere [\onlinecite{MeryPRA2021}]. We thus only provide here a brief description of the method and of the experimental conditions. The 135~keV Ar$^{9+}$ projectile beam was sent to the experiment through the ARIBE/GANIL facility. After beam collimation by a 600~$\mu$m aperture, the projectile ions collided with a supersonic gas jet target composed of $\sim99\%$ of CO molecules (monomers) and $\sim1\%$ of (CO)$_2$ dimers. The scattered projectile were charge selected using a set of deflection plates combined with a position sensitive detector. The fragments of the ionized targets were collected on a second 80~mm diameter position sensitive detector thanks to a 40~V/cm electric field. The position and time of flight (TOF) of each fragment allowed then the identification of their mass-over-charge ratio and the 3-D reconstruction of their momentum in the center of mass of the molecular complex. As reported in [\onlinecite{ MeryPRA2021}], special care has been devoted to the clean selection of the different dissociation channels, to the calibration of the kinetic energy released in the fragmentation and to the optimization of the energy resolution. 

%%%%%%%%%%%%%%%%%%%%%%%%%%%%%%%%%%%%%%%%%%%%
%%%%%%%%%%%%%%%%%%%%%%%%%%%%%%%%%%%%%%%%%%%%
%%%%%%%%%%%%%%%%%%%%%%%%%%%%%%%%%%%%%%%%%%%%
\section{Results and discussion}
Three different fragmentation channels will be discussed.
On one hand, the removal of three electrons from a (CO)$_2$ dimer target can lead to the 2-body dissociation channel, $\rm (CO)_2^{3+} \rightarrow CO^{2+} + CO^+$, or to the 3-body one, $\rm (CO)_2^{3+} \rightarrow CO^+ + C^+ + O^+$. From both channels, information on the lifetime of the states of the transient $\rm CO^{2+}$ dication can be extracted. On the other hand, the fragmentation of isolated $\rm CO^{2+}$ dications resulting from collisions with monomers, $\rm CO^{2+} \rightarrow C^{+} + O^+$, was conjointly studied to serve as a reference for momentum calibration and to be compared with the dissociation of $\rm CO^{2+}$ observed within a $\rm (CO)_2$ dimer target.

%%%%%%%%%%%%%%%%%%%%%%%%%%%%%%%%%%%%%%%%%%%%
%%%%%%%%%%%%%%%%%%%%%%%%%%%%%%%%%%%%%%%%%%%%
\subsection{Experimental results and access to the CO$^{2+}$ lifetime}

The 2-body $\rm (CO)_2^{3+} \rightarrow CO^{2+} + CO^+$ dissociation channel was found to contribute to about 7\% of the relaxation of triply charged $\rm (CO)_2^{3+}$ dimers. Since the TOF of the $\rm CO^{2+}$ molecular ion in the spectrometer is about 2.7~$\mu$s, this channel arises from the population of metastable states with lifetimes that are larger or comparable. 
Its low relative intensity results mainly from the low proportion of such metastable states of the $\rm CO^{2+}$ dication in comparison with states of shorter lifetime leading to the competing 3-body channel $\rm (CO)_2^{3+} \rightarrow CO^+ + C^+ + O^+$. The experimental observables do not allow a direct identification of these very long-lived states that can most probably be assigned to the lowest vibration levels ($\rm \upsilon = 0, 1$) of the $\rm ^3\Pi$ and $\rm ^1\Sigma^+$ state of the $\rm CO^{2+}$ dication [\onlinecite{MrugalaJCP2008}].

\begin{figure}[h!]
\includegraphics[width = 1.\columnwidth]{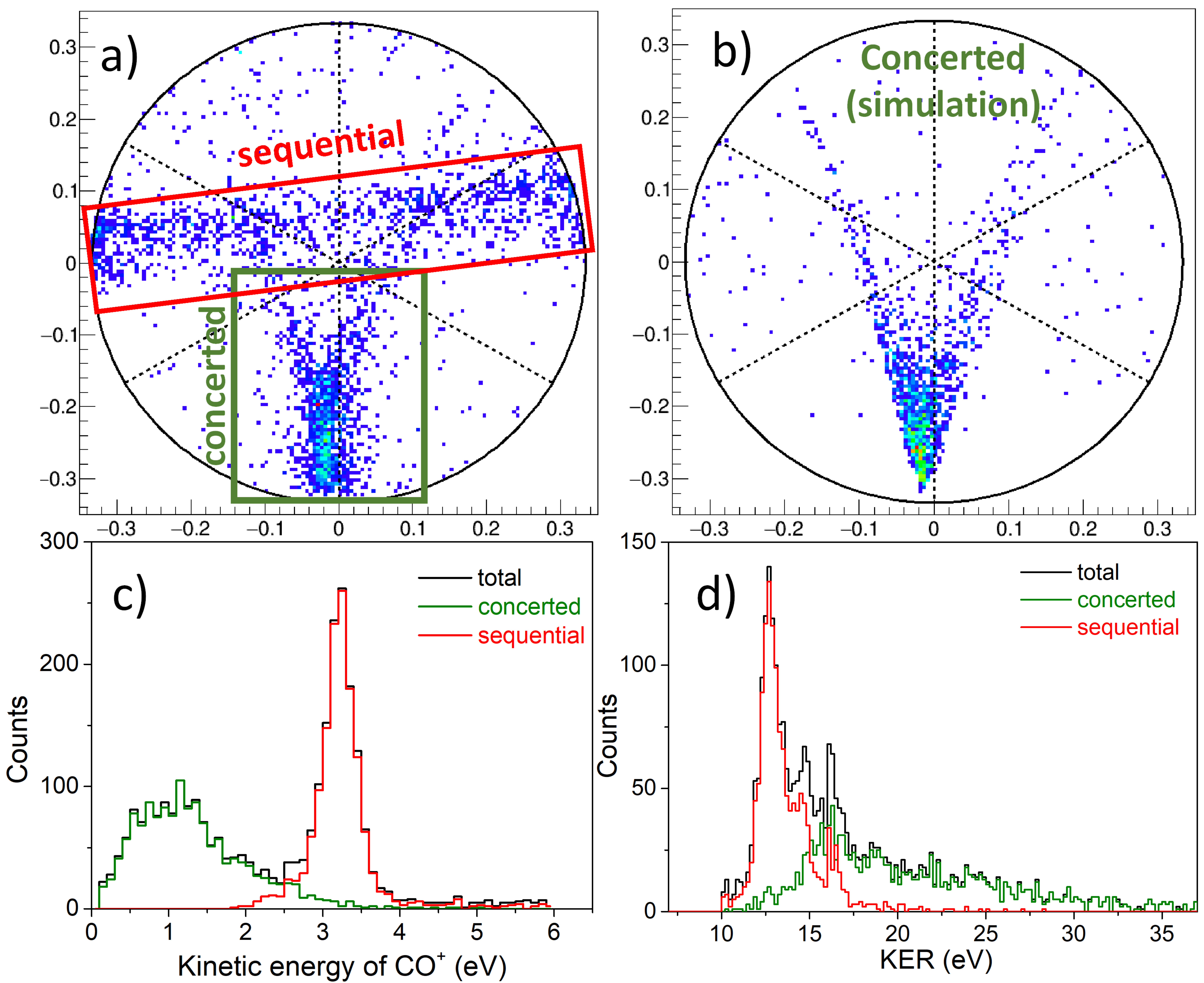}
\caption{\label{fig:Dalitz_E_CO}{Identification of the concerted and sequential fragmentation processes in the $\rm (CO)_2^{3+} \rightarrow CO^+ + C^+ + O^+$ dissociation channel. Panel a) : Dalitz diagram of experimental data indicating the selection windows for concerted and sequential dissociation. Panel b) : Dalitz diagram obtained by numerical simulation for concerted fragmentation events [\onlinecite{ MeryPRA2021}]. Panel c) : kinetic energy of the $\rm CO^+$ molecular ion for concerted (green or light gray) and sequential (red or gray) events. Panel d) : total KER of the 3-body fragmentation for concerted (green or light gray) and sequential (red or gray) events.}}
\end{figure}

The 3-body $\rm (CO)_2^{3+} \rightarrow CO^+ + C^+ + O^+$ channel is more interesting as the sharing of the kinetic energy between the three fragments carries information on both the lifetime and the initial state of the CO$^{2+}$ dication.  For $\rm (CO)_2^{3+}$ [\onlinecite{DingPRL2017}] and $\rm (CO_2)_2^{3+}$ [\onlinecite{SongPRA2019}], a detailed analysis of the 3-body fragmentation dynamics was already performed using a momentum selection of the singly charged molecular ion combined with a Newton representation of the fragments momenta. We use here a slightly different method based on Dalitz plots which provides a more direct selection of different typologies of fragmentation dynamics.
The Fig.~\ref{fig:Dalitz_E_CO}.a shows the experimental Dalitz diagram of the 3-body dissociation of the dimer. This two-dimensional graph represents the reduced squared momenta of each fragment $\rm \varepsilon_i=\frac{p_i^2}{\sum _{i=1}^{3} p_i^2}$ (where the subscript \textit{i} stands for the $\rm C^+$, $\rm O^+$ and $\rm CO^+$ recoiling ions) along the three axis of the diagram (shown as dotted lines). The corresponding Cartesian coordinates are calculated as : $\rm x_D=\frac{\varepsilon_1 - \varepsilon_2}{\sqrt{3}}$ for the horizontal scale and $\rm y_D=\varepsilon_3 - \frac{1}{3}$ for the vertical scale, where \textit{i} is the number of the fragment sorted by ascending order of mass-to-charge ratio.

%%%%%%%%%%%%%
%%%%%%%%%%%%%

Several features are observed in Fig.~\ref{fig:Dalitz_E_CO}.a:  for $\rm y_D < 0$, a dense region at the bottom of the vertical axis followed  by a V-shape tail and for $\rm y_D > 0$, a quasi-horizontal band. Events with $\rm y_D < 0$ correspond to a simultaneous break-up of both the covalent and van der Waals bonds and will be referred as concerted fragmentation events. 
The exact distribution of these events is directly correlated to the dissociation dynamics of the $\rm CO^{2+}$ dication but also to the initial structure of the dimer. In a previous study [\onlinecite{ MeryPRA2021}] assuming a simultaneous 3-body break-up, Monte Carlo (MC) simulations have shown that the lowest part of the Dalitz with its V-shape distribution can be almost perfectly reproduced by the prompt 3-body fragmentation of a CO dimer whose initial angles between the van der Waals bond and the covalent bonds are randomly distributed. This quasi isotrope orientation was found as resulting from the low energy barrier between several conformers of the dimer and from the finite temperature of the gas jet. The Dalitz plot obtained in [\onlinecite{ MeryPRA2021}] by MC simulation for this fast fragmentation process is shown in Fig.~\ref{fig:Dalitz_E_CO}.b for comparison.

Contrarily, the quasi-horizontal band in the upper part of the graph is associated to a sequential fragmentation. In a first step, Coulomb repulsion within the triply ionized dimer leads to the fast break-up of the van der Waals bond. This break-up is then followed by a delayed dissociation of the metastable $\rm CO^{2+}$ dication occurring when both covalent molecular ions have reached their asymptotic kinetic energy. Note that these events can only be observed if the fragmentation occurs at a shorter time than the  2.7~$\mu$s TOF of the CO$^{2+}$ ions. The density distribution of the sequential events is localized close to the edge of the Dalitz graph. This is the signature of an in-plane dissociation: before fragmenting, the $\rm (CO)^{2+}$ rotates in the initial plane containing the center-of-mass of the $\rm CO^+$ ion and the axis of the $\rm (CO)^{2+}$ dication. Such a rotation in the initial plane has recently been reported in the sequential 3-body fragmentation of triatomic molecules [\onlinecite{RajputPRL2018}]. Here we observe a very similar behavior in the two-step dissociation of non-covalent molecular dimers.

Fig.~\ref{fig:Dalitz_E_CO}.c shows the kinetic energy of the $\rm CO^+$ ion for the selection windows of the Dalitz plot corresponding to sequential and concerted events. As previously observed for the fragmentation of $\rm (CO)_2^{3+}$ [\onlinecite{DingPRL2017}] and of $\rm (CO_2)_2^{3+}$ [\onlinecite{SongPRA2019}] dimers, the distributions associated to a concerted or to a sequential dissociation are clearly different. 
The energy of the main peak at 3.2 eV (in red) corresponds to half the Coulomb potential energy of the transient dissociating $\rm CO^+ +CO^{2+}$ system and is a clear signature of a sequential fragmentation. In such a two-step fragmentation process, the $\rm CO^{2+}$ dication breaks up only after the two recoiling molecular ions have reached the asymptotic limit for which their relative Coulomb repulsion vanishes. The $\rm CO^+$ ion thus acquires the exact same kinetic energy as in the 2-body channel $\rm (CO)_2^{3+} \rightarrow CO^{2+} + CO^+$ i.e. 3.2~eV [\onlinecite{ MeryPRA2021}]. The existence of metastable states of the $\rm CO^{2+}$ molecular dication is thereby experimentally and easily evidenced thanks to the presence of the second neighbor molecule in the dimer. 
This is in deep contrast with what has been previously obtained using the same experimental technique for $\rm (N_2)_2$ dimers, where no sign of such a 2-step process could be observed [\onlinecite{MeryPRL2017}]. 
The lower part of the $\rm CO^+$ energy spectrum (green line) corresponds to the concerted fragmentation where the $\rm CO^{2+}$ dication dissociates within a much shorter timescale. In this case, the $\rm CO^+$ acquires a lower kinetic energy because of the fast repulsion of the two $\rm C^+$ and $\rm O^+$ atomic ions following the dissociation of the $\rm CO^{2+}$ dication.
The total KER spectrum in Fig.~\ref{fig:Dalitz_E_CO}.d shows that the sequential fragmentation leads to lower KER than the concerted fragmentation. This can be qualitatively interpreted as resulting from the population of higher excited states of the  $\rm CO^{2+}$ dication for the concerted fragmentation events than for the sequential ones.

Three processes have thus been identified: a concerted fragmentation process corresponding to the population of dissociative or very short life states of the $\rm CO^{2+}$ dication, a sequencial fragmentation process corresponding to the population of metastable states with lifetimes significantly shorter than the 2.7~$\mu$s, and the population of metastable states with lifetimes larger than 2.7~$\mu$s whose dissociation could not be directly observed experimentally.

%For the V-shape selection (green line), the kinetic energy acquired by the $\rm CO^+$ ion is larger than for the other concerted fragmentation events. This can be explained by a different structure of the dimer with either the C$^+$ or O$^+$ ion being emitted in a direction close to the CO$^+$ molecular ion.
%%%%%%%%%%%%%%%%%%%%%%%%%%%%%%%%%%%%%%%%%%%%
%%%%%%%%%%%%%%%%%%%%%%%%%%%%%%%%%%%%%%%%%%%%
\subsection{Simulations}
In order to quantitatively determine the range of lifetimes associated to the sequential process, numerical simulations based on a simple classical model have been performed. This model was used beforehand to determine the geometry of (CO)$_2$ dimers using the concerted fragmentation events and is described in detail in [\onlinecite{ MeryPRA2021}]. 
Briefly, the triply ionized dimer is initially represented as three fixed in space and point-like particles with the respective charges and masses of CO$^+$, C$^+$ and O$^+$, where the C$^+$ and O$^+$ ions compose the dissociating CO$^{2+}$ dication with a bond length of 1.13~{\AA}.
Following the conclusions obtained in [\onlinecite{ MeryPRA2021}], the mean intermolecular distance between the two covalent molecules is set to R=4.2~{\AA}. 
The mean angle between the dimer axis and the CO$^{2+}$ molecular axis is set to ${\theta}_i$=90${\degree}$ where ${\theta}_i$ gives the direction of the carbon atom and ${\theta}_i + 180{\degree}$ the direction of the oxygen atom. Although the initial orientation of the CO$^{2+}$ in respect with the dimer axis is expected to be isotropic [\onlinecite{ MeryPRA2021}], the angle ${\theta}_i$ was here fixed at 90${\degree}$ to ease the interpretation of the simulation. We will show later that this simplification does not change the overall conclusions.
As in [\onlinecite{ MeryPRA2021}], the simulation also accounts for jitters on the mean value of R.

For the first step of the fragmentation, we only consider a pure Coulombic 2-body fragmentation of the dimer and compute the trajectories of two point-like molecular ions CO$^{2+}$ and CO$^+$, up to a time $t$.  The duration $t$ of the first step depends on the chosen decay constant ${\tau}$ of the CO$^{2+}$ dication and follows a probability distribution $P(t)= {\alpha} e^{-t/{\tau}}$. The equations of motion are integrated using a standard Runge-Kutta method and assuming a pure Coulomb repulsion between the two molecular ions. 

\begin{figure}[h!]
\includegraphics[width = 1.\columnwidth]{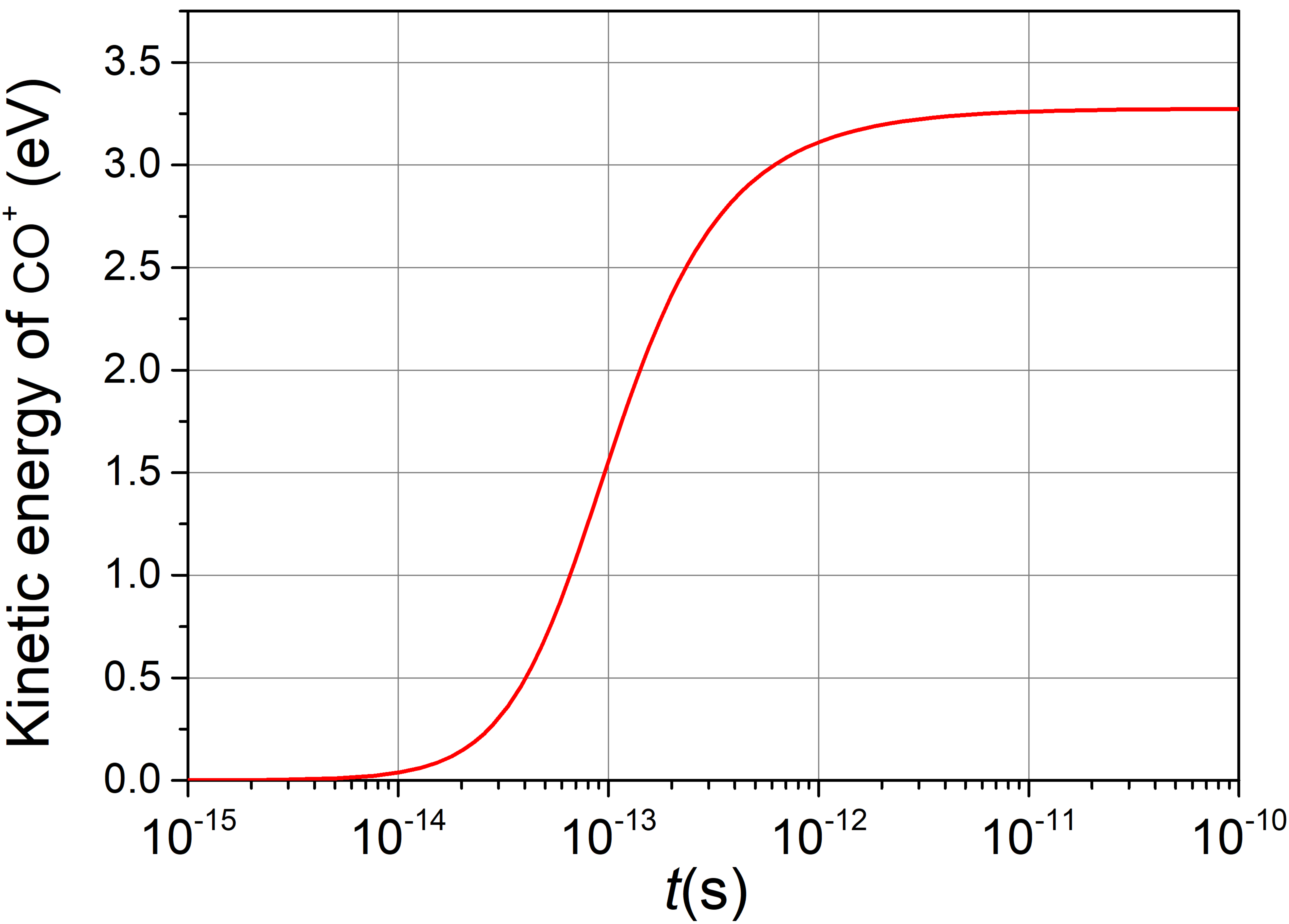}
\caption{\label{fig:FigKECOt}{Evolution of the mean kinetic energy of the CO$^+$ molecular ion as a function of time $t$ in the first step of the sequential fragmentation: 2-body dissociation $\rm (CO)_2^{3+} \rightarrow CO^{2+} + CO^+$}}
\end{figure}

The subsequent dissociation of the CO$^{2+}$ dication is then simulated in a similar way but with a 3-body interaction between the C$^+$, O$^+$ and CO$^+$ ions. 
In this second step, the CO$^{2+}$ dication dissociates at a distance $d$ from the CO$^+$ molecular ion which depends on the decay time $t$ and with a new orientation angle ${\theta}_f$ to account for the free rotation of the CO$^{2+}$ molecular ion prior dissociation. The angle ${\theta}_f$ between the initial dimer axis and the CO$^{2+}$ molecular ion is simply given by ${\theta}_f={\theta}_i + 2{\pi}t/T_{rot}$, where $T_{rot}$ is the rotation period of the CO$^{2+}$ molecule. The value of $T_{rot}$ was set to 0.98~ps, which corresponds to the rotation period obtained within a classical calculation by solving the equation of motion of the first step, approximating the CO$^{2+}$ molecular ion by a C$^+$ ion and a O$^+$ ion linked by a 1.13~{\AA} long rigid bound. Within this approximation, the CO$^{2+}$ rotation takes place in the initial plane containing both the center-of-mass of the $\rm CO^+$ ion and the axis of the $\rm (CO)^{2+}$ dication and is due to the lower mass of the carbon compared to oxygen, leading to a stronger acceleration of the carbon site.

%Two sets of events were simulated: one for the concerted fragmentation process and a second for the sequential fragmentation process. These two sets differ by the final KER distributions associated to each process, as shown in Fig.~\ref{fig:Dalitz_E_CO}.d.  To reproduce the experimental KER distribution obtained for either the concerted or the sequential fragmentation events, a non purely Coulombic repulsion between the C$^+$ and O$^+$ ions from the dissociating CO$^{2+}$ dication is accounted for by a scaling of the repulsive force [\onlinecite{ MeryPRA2021}].
%Because of their very small contribution, the V-shape events were not included in these simulations.

As  shown in Fig.~\ref{fig:Dalitz_E_CO}.d, the concerted fragmentation and the sequential fragmentation differ by the final KER distributions associated to each process. To reproduce in the simulation the KER distribution expected for sequential fragmentation events, a non purely Coulombic repulsion between the C$^+$ and O$^+$ ions from the dissociating CO$^{2+}$ dication is accounted for by a scaling of the repulsive force [\onlinecite{ MeryPRA2021}]. This non Coulombic repulsion is directly inferred from the experimental KER spectrum for sequential events.

\begin{figure}[h!]
\includegraphics[width = 1.\columnwidth]{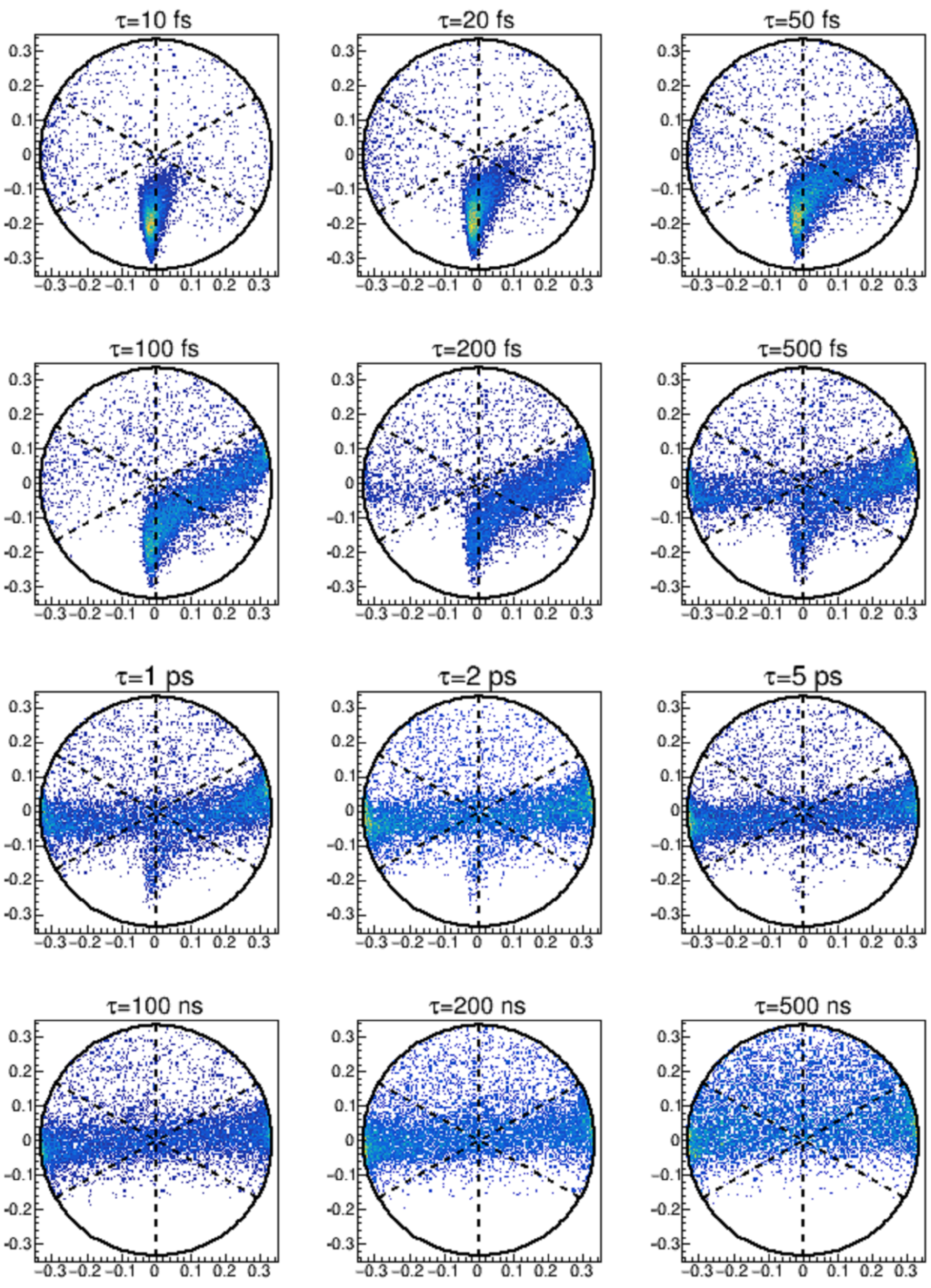}
\caption{\label{fig:DalitzSequential}{Dalitz plots obtained with the Monte Carlo simulations of the sequential fragmentation process for different lifetimes ${\tau}$ of the CO$^{2+}$ dication.}}
\end{figure}

The evolution of the mean kinetic energy of the CO$^+$ molecular ion as a function of the time $t$ during the first step is plotted in Fig.~\ref{fig:FigKECOt} for the sequential fragmentation. The asymptotic energy of 3.2~eV corresponding to the peak observed in Fig.~\ref{fig:Dalitz_E_CO}.c is reached for $t$ values ranging from approximately 1~ps to 10~ps, which indicates that sequential fragmentation events correspond to a fragmentation time $t$ larger than 1~ps. A clearer picture can be obtained by looking at the simulations Dalitz plots shown in Fig.~\ref{fig:DalitzSequential}.  For ${\tau}<$2~ps, one can clearly see the Dalitz distribution moving first towards higher values of $\rm y_D$ which corresponds to higher transfer momentum to the CO$^+$ ion. Then, it expands  to the right (higher momentum transfer to the O$^+$ ion) due to the rotation of the CO$^{2+}$ molecular ion. For lifetimes larger than 500~fs, the angle ${\theta}_f$ reaches 270{\degree} and the left part of the Dalitz (higher momentum transfer to the C$^+$ ion) is also filled, forming the quasi-horizontal band observed in the experimental data. Within this time sequence between 50 fs and 1~ps, the fact that the right side of the horizontal band is filled before the left side arises from our choice to fix ${\theta}_i$ at 90${\degree}$. Starting with an isotropic orientation of the CO$^{2+}$ molecular ion within the dimer would simply result in the simultaneous filling of both sides of the band.
The quasi-horizontal band is finally clearly obtained without deformation (no remaining events for $\rm y_D < 0$) for dissociation lifetimes larger than $\sim$2~ps and remains unchanged up to $\sim$200~ns. 
For large dissociation times $t$, the reconstruction of the fragments momenta should start to fail, the duration of the first step being neglected in the data analysis procedure. The present simulations show that for times up to 200~ns, the reconstruction of the momentum vectors of the C$^+$ and O$^+$ ions remains accurate enough to provide the same pattern on the Dalitz plot as for a 2~fs dissociation time. 
%One should note however that for dissociation times as long as 200~ns, the total KER can be underestimated by a few percent as the real flight times of the C$^+$ and O$^+$ ions after dissociation are 200~ns shorter than accounted for in the momentum reconstruction algorithm.
 For dissociation times larger than 200~ns, the quasi-horizontal band becomes thicker and the upper part of the Dalitz plot, above the band, start to be filled with events. This last change in the Dalitz figure arises when the CO$^{2+}$ fragmentation occurs at a distance $d$ that is sufficiently far away from the collision region to prevent a proper reconstruction of the fragment momenta.
The sequential process observed experimentally in Fig.~\ref{fig:Dalitz_E_CO}.a can thus be attributed to the population of metastable states of the CO$^{2+}$ molecular ion with lifetimes comprised between about 2~ps and 200~ns. 
The same simulations have been performed using the concerted fragmentation KER distribution. They have shown that for events identified with the concerted fragmentation selection window of Fig.~\ref{fig:Dalitz_E_CO}.a, the van der Waals and covalent bonds break appart within less than 10~fs. As can be seen in Fig.~\ref{fig:FigKECOt}, this also corresponds to times that are too short to allow a significant acceleration of the CO$^{+}$ molecular ion.

We did not observe here metastable states with lifetimes of the order 100~fs, but the technique could be particularly suitable to investigate such states.  If the CO$^{2+}$ fragmentation takes place for $t\sim 100$~fs, the two molecular ions are not sufficiently far away to reach the asymptotic part of the Fig.~\ref{fig:FigKECOt} and the resulting Dalitz plots would show specific patterns as in Fig.~\ref{fig:DalitzSequential} for $t$ ranging from 50~fs to 500~fs. The detailed study of such Dalitz plots could then give access to a more precise estimate of the lifetime. Note that the present range of 50~fs to 500~fs is specific to the present fragmentation system. For molecular dimers of different internuclear/molecular distances and fragment ions of different masses, this range would be shifted. Another specificity of the molecular system investigated here is the initial quasi-isotropic angular distribution of the CO molecules within the dimer, which prevents the use of information carried through fragment angular correlations. With a well defined initial orientation of the molecules, subrotational lifetimes of the molecular ions could also be accessed more directly using the so-called native frames approach, as recently demonstrated for SO$^{2+}$ ions resulting from SO$_{2}^{3+}$ ions fragmentation[\onlinecite{RajputNatSR2020}].

%%%%%%%%%%%%%%%%
%%%%%%%%%%%%%%%%

%Equivalent simulations for the concerted fragmentation are shown in Fig.~\ref{fig:DalitzConcerted}. Up to lifetimes  ${\tau}{\leq}$5~fs, the simulations reproduce fairly well the experimental Dalitz plot. For  ${\tau}=10$~fs, a tail that is not observed in the experimental data appears on the upper right side. An upper limit of ${\tau}<10$~fs can thus be deduced for the concerted process.

%%%%%%%%%%%%%%%%
%\begin{figure}[h!]
%\includegraphics[width = 1.\columnwidth]{FigDalitzConcerted.pdf}
%\caption{\label{fig:DalitzConcerted}{Dalitz plots obtained with the Monte Carlo simulations of the concerted fragmentation process for different lifetimes ${\tau}$ of the CO$^{2+}$ dication.}}
%\end{figure}
%%%%%%%%%%%%%%%%

%%%%%%%%%%%%%%%%%%%%%%%%%%%%%%%%%%%%%%%%%%%%
%%%%%%%%%%%%%%%%%%%%%%%%%%%%%%%%%%%%%%%%%%%%
\subsection{Identification of the metastable states}
\begin{figure}[h!]
\includegraphics[width = 1.\columnwidth]{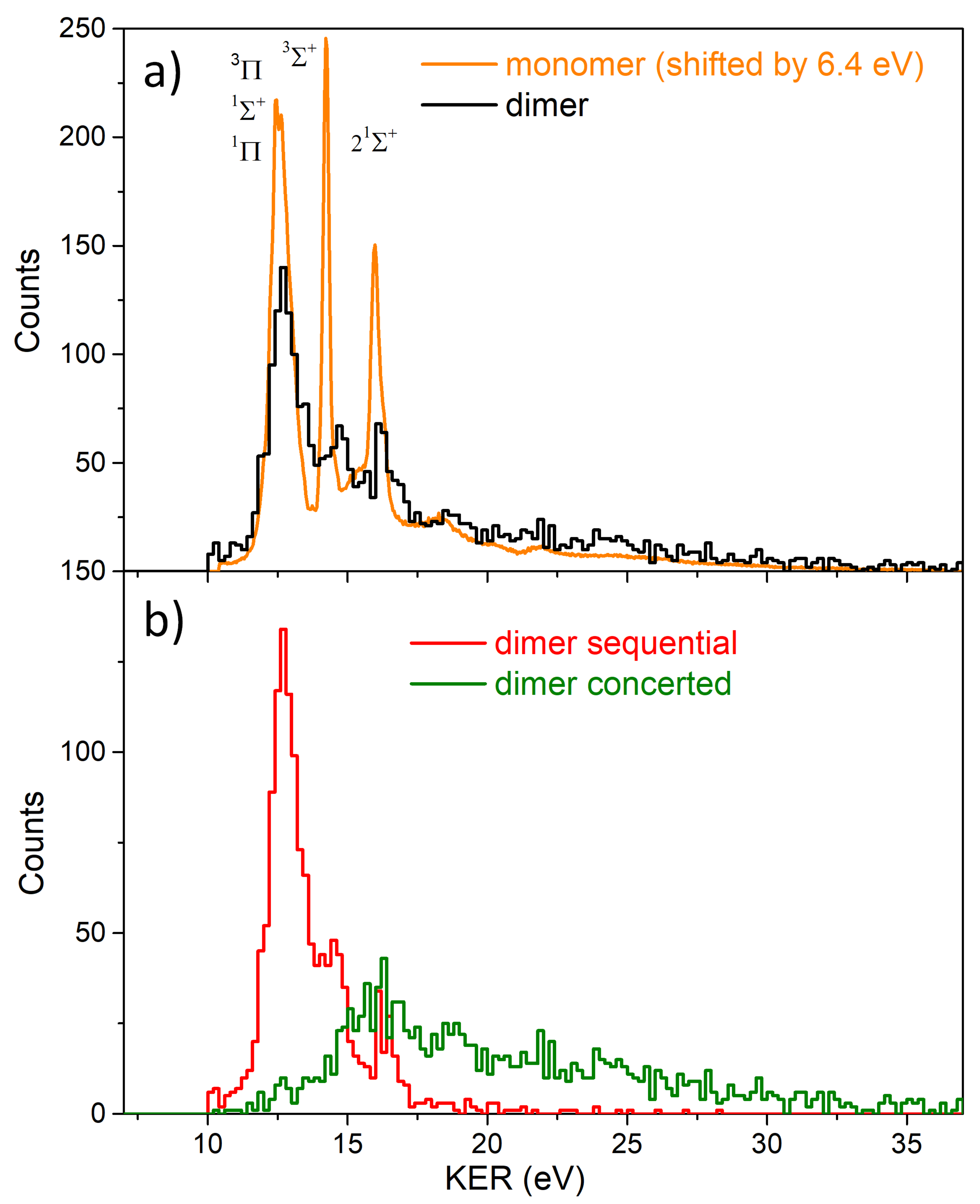}
\caption{\label{fig:KER_3bodie}{Upper panel : Total KER spectrum for the $\rm (CO)_2^{3+} \rightarrow CO^+ + C^+ + O^+$ fragmentation channel (black line) and KER spectrum of the $\rm CO^{2+} \rightarrow C^+ + O^+$ fragmentation of monomer targets shifted by 6.4~eV (orange or light gray line). The KER spectrum obtained for monomers is normalized to the one obtained with dimers. Lower panel : contributions of the sequential fragmentation (red or gray line) and concerted fragmentation events (green or light gray line) after a data selection performed using the selection windows of Fig.~\ref{fig:Dalitz_E_CO}.a. }}
\end{figure}

As already described in a previous study on nitrogen dimers $\rm (N_2)_2$ [\onlinecite{MeryPRL2017}], the total KER of the 3-body channels (sum of the kinetic energy of the three emitted fragments) can be compared to the KER spectrum of the dication dissociation from monomer targets by simply accounting for the extra kinetic energy resulting from the pure Coulomb repulsion with the neighbor molecular ion. Such a comparison is provided in the upper panel of Fig.~\ref{fig:KER_3bodie} for $\rm (CO)_2$ dimers. For the 3-body dissociation channel, assuming Re=4.2~{\AA} and a pure Coulomb model, the expected energy shift is $\rm \Delta KER=6.4~eV$ and agrees perfectly with our measurement. 

%%%%%%%%%%%%%%%%
%%%%%%%%%%%%%%%%

The global shape of the total $\rm KER$ spectrum (black line) is quite similar to the KER distribution obtained with the monomer CO target (orange line). Three peaks at low energy are clearly visible for the dimer, although, they are not as well resolved as for the monomer. This larger width of the peaks can be explained by the additional jitter in total KER due to the vibrational motion of the dimer bond. Following the identification from [\onlinecite{Lundqvist1995}], the first peak at 6.2 eV corresponds to the population of the $\rm ^3\Pi$, $\rm ^1\Sigma^+$ and $\rm ^1\Pi$ states, the second one at 7.8~eV to the $\rm ^3\Sigma^+$ state and the third one at 9.6~eV has been assigned to the $\rm 2 ^1\Sigma^+$ state. Other high lying states leading to a wide KER distribution between 17~eV and 35~eV are also populated for both the dimer and the monomer targets. 
Small differences can nevertheless be observed between dimer and monomer targets when looking closely at the relative population of the different molecular states. With dimer targets, KER higher than 17~eV are favored while lower KER values (corresponding in particular to the $\rm ^3\Sigma^+$ and $\rm 2 ^1\Sigma^+$ states) are disfavored. As both collision systems are investigated simultaneously with the same setup and thus with the same acceptance, this is a signature of a difference mainly resulting from the primary collision process. The removal of three electrons from the dimer target (two from one of the CO molecule and one from its neighbor) requires smaller impact parameters than the removal of two electrons from a single CO molecule. In the case of dimer targets, this leads to a larger population of high excitation states (from 17~eV to 35~eV) and to a smaller population of the ground and low lying states corresponding to the three peaks in KER lying between 10~eV and 17~eV).
 
 It also clearly appears (lower panel of Fig.~\ref{fig:KER_3bodie}) that the sequential fragmentation is only associated to the lower energy part of the KER spectrum whereas concerted fragmentation occurs mainly for higher KER values. 
 
For the events corresponding to a sequential fragmentation, the dissociation of $\rm CO^{2+}$ occurs far enough from the $\rm CO^{+}$ partner to be directly compared with the dissociation of a $\rm CO^{2+}$ dication originating from a monomer. For this purpose, the sum of the kinetic energy of the $\rm C^+$ and $\rm O^+$ ions can be calculated in the frame of the center-of-mass of the transient recoiling $\rm CO^{2+}$ molecular ion. As the momentum of the $\rm CO^{2+}$ and $\rm CO^+$ are equal but have opposite directions, the corresponding momentum vectors in the recoiling center-of-mass frame are: 
	\begin{center}
	$\rm \overrightarrow{p}^*_{C^+}=\overrightarrow{p}_{C^+}+\frac{m_C}{m_{CO}}\overrightarrow{p}_{CO^+}$
	\end{center}
	\begin{center}
	$\rm \overrightarrow{p}^*_{O^+}=\overrightarrow{p}_{O^+}+\frac{m_O}{m_{CO}}\overrightarrow{p}_{CO^+}$
	\end{center}
The total kinetic energy of the two atomic ions in this frame is given by:
	\begin{center}
	$\rm KE^*=\frac{\left\lVert\overrightarrow{p}^*_{C^+}\right\rVert^2}{2 m_C}+\frac{\left\lVert\overrightarrow{p}^*_{O^+}\right\rVert^2}{2 m_O}$
	\end{center}

%%%%%%%%%%%%%%%	
\begin{figure}[h!]
\includegraphics[width = 1.\columnwidth]{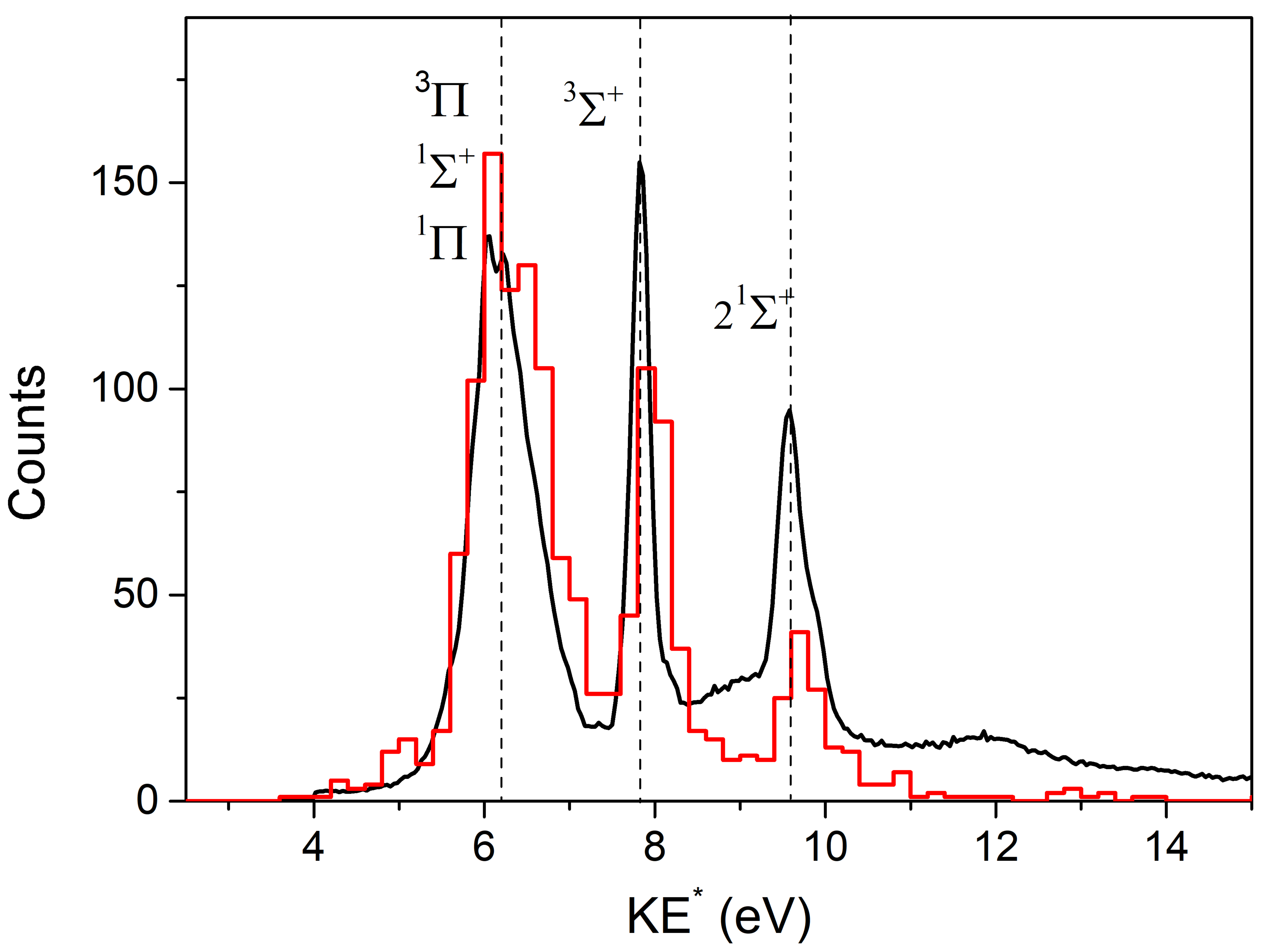}
\caption{\label{fig:KE12_cm}{Sum of the kinetic energy of the $\rm C^+$ and $\rm O^+$ ions calculated in the center-of-mass of the CO$^{2+}$ transient ion for the sequential fragmentation event selection (red or gray line). KER spectrum resulting from collisions with monomer targets (black line).The KER spectrum obtained for monomers is normalized to the one obtained with dimers in the range between 4 and 11~eV. One bin correspond to 0.2~eV on the spectrum obtained with dimers.}}
\end{figure}
%%%%%%%%%%%%%%%%

The $\rm KE^*$ spectrum for the sequential fragmentation events is shown in Fig.~\ref{fig:KE12_cm} along with the KER spectrum obtained with CO monomers. The use of the $\rm KE^*$ observable allows a clearer comparison with the monomer as it directly provides the kinetic energy released in the dissociation of the CO$^{2+}$ transient dication. Note that this observable is only meaningful for sequential fragmentation events: if the CO$^{2+}$ dissociation occurs too close to the neighbor CO$^+$ ion, the dissociation must be treated as a real 3-body process. The peaks in KE* displayed in Fig.~\ref{fig:KE12_cm} are now well resolved because the KE* spectrum does not depend on the initial dimer bond length (asymptotic energy of the CO+ ion). Fig.~\ref{fig:KE12_cm} allows to confirm the previous identification of the populated states. The peaks respective mean positions are very close to the ones obtained with the monomer target, but for the $\rm ^3\Sigma^+$ and $\rm 2 ^1\Sigma^+$ states, they are clearly shifted by $\sim$0.1-0.2~eV. Again, as both collision systems with dimer and monomer targets are investigated simultaneously in a single experiment and as we use the same analysis procedure (with identical calibration), this shift cannot be imputed to an experimental bias. Such a shift has been observed in a previous work [\onlinecite{DingPRL2017}] and was attributed to the fact that within the first step of the fragmentation, when the van der Waals bond is broken, a small part of the energy is transferred to the CO$^{2+}$ dication to populate higher vibrational states. The KER obtained for the monomer indicates that for the $\rm ^3\Sigma^+$ and $\rm 2 ^1\Sigma^+$ states, the lowest vibrational level $\rm \upsilon =$~0 is dominantly populated. The shift of $\sim$0.2~eV observed for these states when using dimer targets can thereby be explained by a population transfer due to the van der Waals bond break-up from the $\rm \upsilon =$~0 levels towards the $\rm \upsilon =$~1 levels lying at respectively 0.26~eV and 0.29~eV higher energy.

\begin{table}
\begin{ruledtabular}
\begin{tabular}{ccccc}
{State} 						& {Vib. level}	& {KER (eV)} 									&  \multicolumn{2}{c}{Lifetimes} \\
\textrm{} 					& \textrm{} 		& {[\onlinecite{Lundqvist1995}]}					&{Exp.}										& {Calc. [\onlinecite{MrugalaJCP2008}]} \\
\colrule
\textrm{}					& $0, 1$ 				& \textrm{}  							& $\rm > 10 \mu s$[\onlinecite{PenentPRL1998}] 		& $\rm > 1 s$ \\
\textrm{}	 				& $2$ 				& $5.6$								& $\rm200 ns$[\onlinecite{PenentPRL1998}] 			& $\rm 800 ns$ \\
$\rm ^3\Pi$ 				& $3$ 				& $5.8$								& $\rm <100 ns$[\onlinecite{PenentPRL1998}] 			& $\rm 1 ns$ \\
\textrm{} 					& $4-11$ 				& $\rm 6.0-6.8$							&$\rm 10 fs-50 ns$[\onlinecite{Lundqvist1995}] 	 		& $\rm 10 ps-1 ns$ \\
\textrm{} 					& $\rm \geq12$ 		& \textrm{}							& \textrm{} 		 							& $\rm < 1 ps$ \\
\colrule
\textrm{}					& $0$ 				& \textrm{}  							& $\rm > 10 \mu s$[\onlinecite{PenentPRL1998}] 		& $\rm 20 s$ \\
\textrm{}	 				& $1$ 				& $5.7$								& $\rm 700 ns$[\onlinecite{PenentPRL1998}] 			& $\rm 100 \mu s$ \\
$\rm ^1\Sigma^+$ 			& $2$ 				& $6.0$								& \textrm{} 	 								& $\rm 500 ns$ \\
\textrm{} 					& $3$ 				& \textrm{}							& \textrm{} 	 								& $\rm 80 ns$ \\
\textrm{} 					& $4-23$ 				& \textrm{}							& \textrm{} 	 								& $\rm 10 ps-1 ns$ \\
\textrm{} 					& $\rm \geq24$ 		& \textrm{}							& \textrm{} 		 							& $\rm < 1 ps$ \\
\colrule
\textrm{}					& $0$ 				& $5.8$  								& $\rm 200 ns$[\onlinecite{PenentPRL1998}] 			& $\rm 7 \mu s$ \\
$\rm ^1\Pi$ 				& $1$ 				& $6.0$								& $\rm < 100 ns$[\onlinecite{PenentPRL1998}]			& $\rm 10 ns$ \\
\textrm{} 					& $2-10$ 				& $\rm 6.2-7.5$							&$\rm 10 fs-50 ns$[\onlinecite{Lundqvist1995}]			& $\rm 10 ps-1 ns$ \\
\textrm{} 					& $\rm \geq23$ 		& \textrm{}							& \textrm{} 		 							& $\rm < 1 ps$ \\
\colrule
\textrm{}					& $0$ 				& $7.8$  								&$\rm 10 fs-50 ns$[\onlinecite{Lundqvist1995}]  			& $\rm 10 ns$ \\
$\rm ^3\Sigma^+$	 		& $1$ 				& $8.1$								&$\rm 10 fs-50 ns$[\onlinecite{Lundqvist1995}]  			& $\rm 1 ns$ \\
\textrm{} 					& $2-6$ 				& $\rm >8.3$							& \textrm{} 	 								& $\rm 10 ps-1 ns$ \\
\colrule
$\rm 2^1\Sigma^+$	 		& $0$ 				& $9.5$								&$\rm 10 fs-50 ns$[\onlinecite{Lundqvist1995}] 			& \textrm{} \\
\textrm{} 					& $1$ 				& $9.8$								& $\rm 10 fs-50 ns$[\onlinecite{Lundqvist1995}] 	 		& \textrm{} \\

\end{tabular}
\end{ruledtabular}
\caption{\label{tab:table_dication} 
KER and lifetimes of the ground state and low lying states of the $\rm CO^{2+}$ monomer dication. Theoretical lifetimes are extracted from figure 1 of [\onlinecite{MrugalaJCP2008}].
}
\end{table}

In the present experiment, we identified the population of molecular states of the CO$^{2+}$ dication associated to a sequencial fragmentation of the dimer target: the manyfolds of the $\rm ^3\Pi$, $\rm ^1\Sigma^+$ and $\rm ^1\Pi$ states, and the $\rm \upsilon =$~0 and $\rm \upsilon =$~1 vibrational levels of the $\rm ^3\Sigma^+$ and $\rm 2 ^1\Sigma^+$ states. The analysis of the Dalitz plots of Fig.~\ref{fig:DalitzSequential} for this sequencial process have shown that all these molecular states have lifetimes in the 2~ps-200~ns range, which is in agreement with the previous measurements and calculations provided in table~\ref{tab:table_dication}.

%In the present experiment, we can conclude that the manyfolds of the $\rm ^3\Pi$, $\rm ^1\Sigma^+$ and $\rm ^1\Pi$ states have lifetimes in the 2~ps-200~ns range, which is in agreement with the previous measurements and calculations provided in table~\ref{tab:table_dication}. The $\rm \upsilon =$~0 and $\rm \upsilon =$~1 vibrational levels of the $\rm ^3\Sigma^+$ and $\rm 2 ^1\Sigma^+$ states are also found with lifetimes in the same range, leading to sequential fragmentation. For the $\rm ^3\Sigma^+$ state, this measurement is also consistent with previous calculations leading to lifetimes of 10~ns and 1~ns for the vibrational levels  $\rm \upsilon =$~0 and $\rm \upsilon =$~1, respectively.
%Up to our knowledge, there was no prior lifetime estimation for the third peak at 9.6 eV, assigned to the $\rm 2 ^1\Sigma^+$ state [\onlinecite{Lundqvist1995}]. We show here that the  two vibrational levels $\rm \upsilon =$~0 and $\rm \upsilon =$~1 from this electronic state also contributes to the sequential process and thus reveal a significant contribution of dissociation times in the 2 ps to 200 ns range.

%%%%%%%%%%%%%%%%%%%%%%%%%%%%%%%%%%%%%%%%%%%%
%%%%%%%%%%%%%%%%%%%%%%%%%%%%%%%%%%%%%%%%%%%%
\subsection{Fast dissociation channels }

Concerted fragmentation occurs only for KER values higher than 12~eV and spreading up to 35~eV (Fig.~\ref{fig:KER_3bodie}.b). These KER values can be attributed to the population of excited states of the $\rm CO^{2+}$ dication with dissociative potential energy curves or with lifetimes shorter than 10~fs, leading to a quasi-simultaneous break-up of the van der Waals and covalent bonds. The lower energy part of the KER spectrum for concerted fragmentation events overlaps with the $\rm 2 ^1\Sigma^+$ state manyfold identified in the sequential selection. 
But the very fast dissociation time observed for this concerted process implies that this part of the KER distribution arises from different initial molecular electronic (or vibrational) states.

It is important to note that the large tail in KER between 17~eV and 35~eV is also observed in the KER spectrum obtained with monomer targets when accounting for the 6.4~eV shift in KER (Fig.~\ref{fig:KER_3bodie}.a). This indicates that such excited states leading to fast molecular dissociation are also populated in the collisions with monomer targets. It is thus found in contradiction with the conclusions of the former work of Ding et al. using short laser pulse ionization [\onlinecite{DingPRL2017}].
In this previous study, the low lying states $\rm ^3\Pi$, $\rm ^1\Sigma^+$ and $\rm ^1\Pi$ of CO$^{2+}$ were dominantly populated both for the dimer and monomer targets. When compared to the KER obtained with monomer targets the $\rm KE^*$ distribution of the sequential fragmentation process was found almost identical. However, the KER distribution obtained with dimers for the fast dissociation process was found to be $\sim$1~eV higher than for the sequential fragmentation. This fast process was then interpreted as resulting from the presence of a neighbor CO$^+$ ion in the dimer inducing a symmetry breaking of the $\rm ^3\Pi$ state. Indeed, their calculations of potential energy curves accounting for the presence of the neighbor CO$^+$ ion revealed a fragmentation pathway involving both an avoided crossing with the dissociative $\rm ^3\Sigma^-$ state and the appearance of a new electronic state, leading to a final KER higher by 1.2~eV. They concluded that the fast dissociation process was $only$ enabled by the presence of the neighbor ion and that this fast process should not be expected with CO$^{2+}$ dications from a monomer target.

The results obtained in our experiment are similar in two aspects: we only observe sequential fragmentation for KER values below 17~eV ($\rm ^3\Pi$, $\rm ^1\Sigma^+$ and $\rm ^1\Pi$ states of CO$^{2+}$), and a higher KER distribution for concerted dissociation than for sequential dissociation. However, the KER spectrum of the present work for concerted fragmentation is much wider (extending up to 35~eV) with a maximum of the distribution peaking at 16.5~eV. This KER distribution cannot be explained by the relaxation process invoked previously in [\onlinecite{DingPRL2017}]. First, the shift in KER between the maxima of the distributions obtained for concerted and sequential processes is close to 4~eV, which is much higher than the value of 1.2~eV predicted in [\onlinecite{DingPRL2017}].
Moreover, the tail between 17~eV and 35~eV obtained for the concerted fragmentation is also observed with monomer targets (Fig.~\ref{fig:KER_3bodie}.a) which demonstrates that the process leading to such KER values is not due to the presence of a neighbor ion. The higher KER values obtained here must thus arise from the direct population of high lying excited states leading to fast dissociation.

Several conclusions can be drawn from these observations.

Collisions with highly charged ions lead to the population of higher excited states of the CO$^{2+}$ dication than with short laser pulses, both for dimer and monomer CO targets. The population of highly excited states could even be enhanced at higher collision energy, as previously observed in [\onlinecite{TarisienJPB2000}].

While the $\rm ^3\Pi$, $\rm ^1\Sigma^+$ and $\rm ^1\Pi$ states of CO$^{2+}$ show lifetimes systematically larger than 2~ps, higher excited states all lead to fast dissociation. 

As such excited states were observed both with dimer and monomer targets, this fast dissociation process is not dependent on the presence of a neighbor ion, and fast dissociation must also occur with monomer CO targets.

Finally, the present observations confirm the conclusions obtained in [\onlinecite{MeryPRL2017}] for the study of the fragmentation of (N$_2$)$_2$ dimers, showing no evidence for a significant effect of the neighbor molecular ion beside the shift in KER due to the Coulomb repulsion.

%%%%%%%%%%%%%%%%%%%%%%%%%%%%%%%%%%%%%%%%%%%%
%%%%%%%%%%%%%%%%%%%%%%%%%%%%%%%%%%%%%%%%%%%%
%%%%%%%%%%%%%%%%%%%%%%%%%%%%%%%%%%%%%%%%%%%%
\section{CONCLUSION}

Beyond providing information on the geometry of van der Waals molecular clusters, COLTRIMS and CEI turn up to be smart tools to probe the lifetimes of the molecular states populated in collisions with highly charged ions. 
For the 3-body fragmentation channel $\rm (CO)_2^{3+} \rightarrow CO^+ + C^+ + O^+$, about half of the events have been assigned to a sequential dissociation resulting from the population of metastable states of the $\rm CO^{2+}$ dication with lifetimes between 2~ps and 200~ns. These states result in low KER values and are associated to low excited states of the dication, namely, the $\rm ^3\Pi$, $\rm ^1\Sigma^+$ and $\rm ^1\Pi$ states, and the $\rm \upsilon =$~0 and $\rm \upsilon =$~1 vibrational levels of the $\rm ^3\Sigma^+$ and $\rm 2 ^1\Sigma^+$ states. In a more general way, this experimental approach really opens new opportunities to get insights into dication molecular states and their associated lifetimes.
The present technique would indeed be particularly well suited for the study of metastable states with lifetimes of the order of 100~fs, where a more precise lifetime estimate could be obtained. With dimer targets produced with low rotational energy and with a well known initial molecular orientation, this technique would also provide access to the measurement of any subrotational lifetime of molecular dications using the native frames representation, as recently performed in [\onlinecite{RajputNatSR2020}]. 

In the present work, the population of higher excited states leading to a prompt dissociation and to a concerted fragmentation of the dimer was also clearly evidenced. The analysis of the KER distribution for this process has shown that the fast dissociation cannot be explained by the effect of the neighbor molecular ion. Contrarily, the similarity between the KER spectra obtained for dimer and monomer targets indicates the population of fast dissociating states in collisions with monomer targets. Altogether, this analysis shows a negligible effect of the neighbor molecular ion in the relaxation process of the (CO)$_2^{3+}$ system.

\section{ACKNOWLEDGMENTS}
The experiment was performed at the Grand Acc\'el\'erateur National d\textquotesingle Ions Lourds (GANIL)  by means of the CIRIL Interdisciplinary Platform, part of CIMAP laboratory, Caen, France. The authors want to thank the CIMAP and GANIL staff for their technical support.

\nocite{*}

\bibliography{CO-dimers_sequential_concerted}% Produces the bibliography via BibTeX.

\end{document}